\documentclass[nofootinbib]{revtex4}
\usepackage[english]{babel}
\usepackage{array,booktabs}   
\usepackage{array} 
\usepackage{amsmath} 
\usepackage{relsize}
\usepackage{wrapfig}
\usepackage{calc}
\usepackage{pdflscape}
\usepackage{color}
\usepackage{float}
\usepackage[font=small,labelfont=bf]{caption}
\usepackage{graphicx}
\usepackage{amsmath}
\usepackage[nodisplayskipstretch]{setspace}

\usepackage{setspace}
\usepackage{tabularx}

\usepackage{float}
\usepackage{color}
\usepackage{amsmath}
\usepackage{float}
\usepackage{calc}
\usepackage{pdflscape}
\usepackage{color}
\usepackage{float}
\usepackage{graphics}
\usepackage{epsfig}
\usepackage{epstopdf}
\usepackage{amssymb}
\usepackage[font=small,labelfont=bf]{caption}
\usepackage{graphicx}
\usepackage{epstopdf}
\usepackage{appendix}
\usepackage{romannum}
\usepackage{soul}
\usepackage{color}

\definecolor{blizzardblue}{rgb}{0.67, 0.9, 0.93}
\definecolor{bubblegum}{rgb}{0.99, 0.76, 0.8}
\usepackage[urlcolor=blizzardblue]{hyperref}
\hypersetup{
    colorlinks=false,
    linkcolor=blue,
    filecolor=magenta,      
    urlcolor=bubblegum,
}
\begin{document} 

\title{Nuclear Effects and CP Sensitivity at DUNE}
\author{ Srishti Nagu$^{1}$ \footnote{E-mail: srishtinagu19@gmail.com},Jaydip Singh$^{1}$ \footnote{E-mail: jaydip.singh@gmail.com}, Jyotsna Singh$^{1}$ \footnote{E-mail: singh.jyotsnalu@gmail.com}, R.B. Singh$^{1}$ \footnote{E-mail: rajendrasinghrb@gmail.com}} 

\affiliation{Department Of Physics, University of Lucknow, Lucknow, India.$^{1}$}

\begin{abstract}
 The precise measurement of neutrino oscillation parameters is one of the highest priorities in neutrino
oscillation physics. To achieve the desired precision, it is necessary to reduce the systematic uncertainties
related to neutrino energy reconstruction. An error in energy reconstruction is propagated to all the oscillation
parameters, hence a careful estimation of neutrino energy is required. To increase the statistics, neutrino oscillation 
experiments use heavy nuclear targets like Argon(Z=18). The use of these nuclear targets introduces nuclear effects that severely impact 
the neutrino energy reconstruction which in turn poses influence in the determination of neutrino oscillation parameters. In this work, we
have tried to quantify the presence of nuclear effects on the bounds of CP phase by DUNE using final state interactions.


\end{abstract}
\maketitle

\section {Introduction}	
Neutrino oscillation physics has entered into the era of precision measurement from the past two decades. Significant achievements have been made in the determination of the known neutrino oscillation parameters and continuous attempts are being made to estimate the unknown neutrino oscillation parameters precisely. The neutrino oscillation parameters governing the three flavor neutrino oscillation physics are three mixing angles $\theta_{12}, \theta_{13}, \theta_{23}$, a leptonic CP phase $\delta_{CP}$ and two mass-squared differences, $\Delta m_{21}^2$ (solar mass splitting) and $\Delta m_{31}^2$ (atmospheric mass splitting). 
Many experiments working in collaboration \cite{global1,global2,Minos,T2K} worldwide have led to the precise determination of the above mentioned neutrino oscillation parameters leaving some ambiguities yet to be resolved. The unknown oscillation parameters in the picture are: (1) the octant of $\theta_{23}$ whether it lies in the lower octant ($\theta_{23}<\pi/4$) or in the higher octant ($\theta_{23}>\pi/4$) (2) the sign of $|\Delta m^{2}_{31}|$ i.e. neutrino mass eigenstates $m_{i}$(i=1,2,3) are arranged in normal order ($m_{1} \ll m_{2} \ll m_{3}$) or inverted order ($m_{2} \approx m_{1} \gg m_{3}$) (3) the leptonic $\delta_{CP}$ phase which can lie in the entire range $-\pi<0<+\pi$. Accurate measurement of the leptonic CP phase can lead to further studies on the origin of leptogenesis \cite{lepto} and baryon asymmetry of the universe \cite{baryon}. Determination of the precise value of $\delta_{CP}$ is also required for explaining the phenomenon of sterile neutrinos \cite{jd}. 
The global analysis results as indicated in \cite{global3} report current bounds on oscillation parameters which have been performed by several experimental groups.

A defining challenge for neutrino experiments is to determine the incoming neutrino energy since the configuration of the outgoing particles and kinematics of the interaction within the nucleus are completely unknown. In collider experiments, the neutrino beams are generated via secondary decay products which assign a broad range of energies to the neutrinos hence their energy reconstruction becomes difficult. Here the neutrino energy is reconstructed from final state particles that are produced in neutrino-nucleon interactions. The present-day neutrino oscillation experiments use heavy nuclear targets like argon(Z=18), in order to collect large event statistics. With a nuclear target, where neutrinos interact with fermi moving nucleons, uncertainties in the initial state particles produced at the primary neutrino-nucleon interaction vertex arise. These nuclear effects are capable enough to change the identities, kinematics and topologies of the outgoing particles via final state interactions (FSI) and thus hiding the information of the particles produced at the initial neutrino-nucleon interaction vertex which gives rise to fake events. Detailed discussion regarding the impact on atmospheric oscillation parameters due to the presence of FSI in the QE interaction process can be found in \cite{coloma} and due to fake events stemming from QE and RES processes can be found in \cite{snaaz}. The impact of cross-sectional uncertainties on the CP violation sensitivity can be found in \cite{srish}. The impact of missing energy on the measurement of the CP violating phase in the $\theta_{23}-\delta_{CP}$ plane has been previously performed in \cite{missencoloma}. For further studies on the impact on neutrino oscillation parameters due to the presence of FSI, one can refer to \cite{nuceffect1,nuceffect2,nuceffect3}.

In this work, we attempt to study the impact of nuclear effects imposed by FSI in the QE(Quasi Elastic), resonance (RES) and deep inelastic scattering (DIS) interaction processes. Understanding nuclear effects will give us a handle to filter out true events from the fake events in a given neutrino-nucleon interaction which will lead to an accurate measurement of neutrino oscillation parameters.

\section{Neutrino Oscillation Studies with the Long-baseline Experiment-DUNE:}
The Deep Underground Neutrino Experiment, DUNE \cite{dune1,dune2}, an upcoming long baseline neutrino oscillation experiment, to be set up in the US is aiming for discovering the unknown oscillation parameters and explore new physics. The 1300−km baseline, stretching from LBNE facility at Fermilab to Sanford Underground Research Facility (SURF) at South Dakota, is ideal for achieving the desired sensitivity for CP violation and mass hierarchy. The far detector will be composed of 40 ktons of liquid argon (nuclear target with Z=18) as detector material which will provide large event statistics. The DUNE-LBNF flux spreads in the energy range 0.5 to 10 GeV, with average energy peaking at 2.5 GeV. It is composed of QE, Resonance, DIS and Coherent neutrino-nucleon interaction processes with resonance being the dominant interaction process in this energy regime. The energy dependent cross-section is different for each interaction process.

To evaluate the sensitivity of LBNE and to optimize the experimental design, it is important to accurately predict the neutrino flux presented in Figure 1(left panel) produced by the neutrino beam. The corresponding neutrino oscillation probability is presented in Figure 1(right panel) for $\nu_{\mu}$ disappearance and $\nu_{e}$ appearance channels.

\begin{figure}
\centering
\includegraphics[scale=.28]{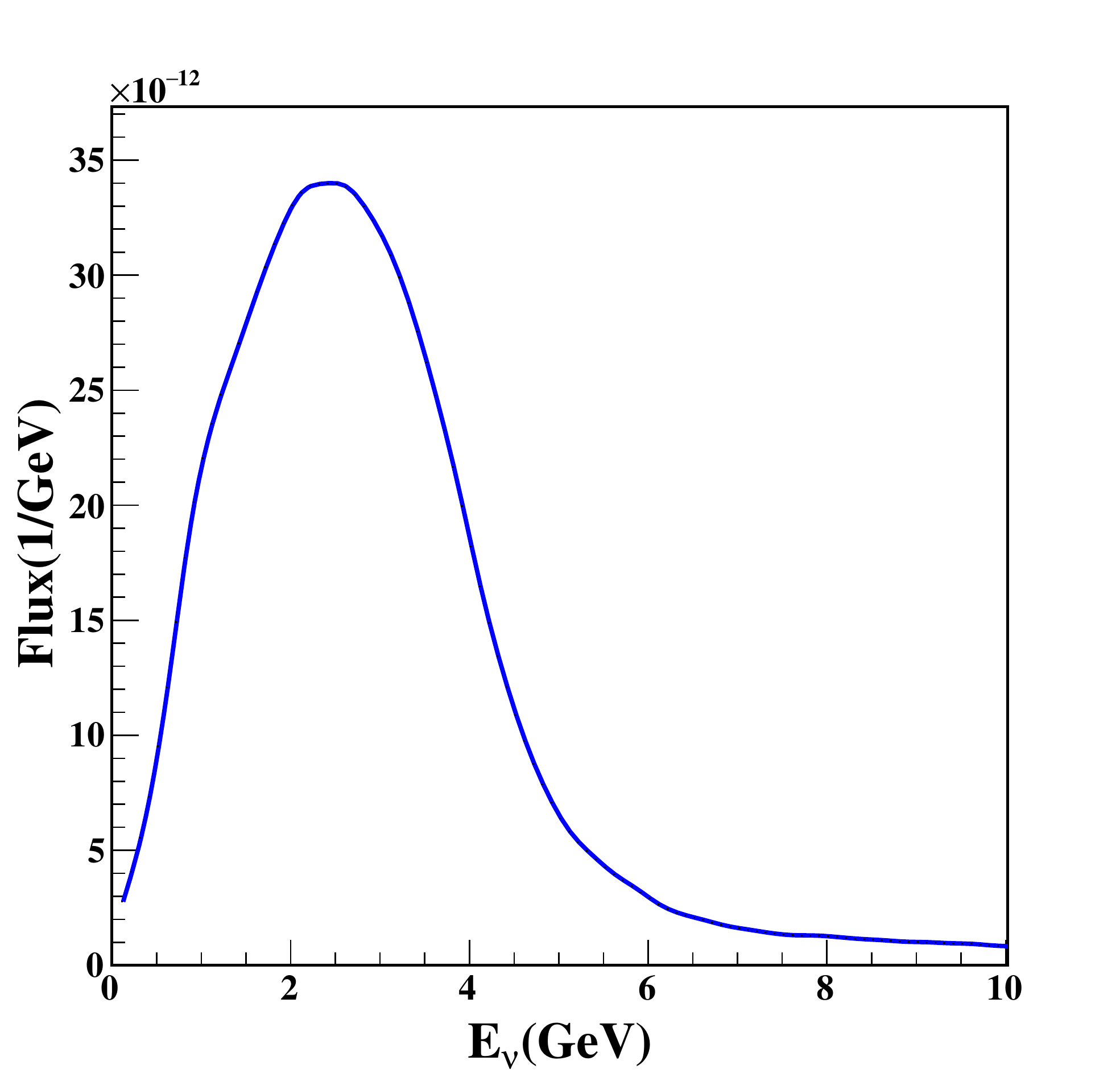}
\includegraphics[scale=.28]{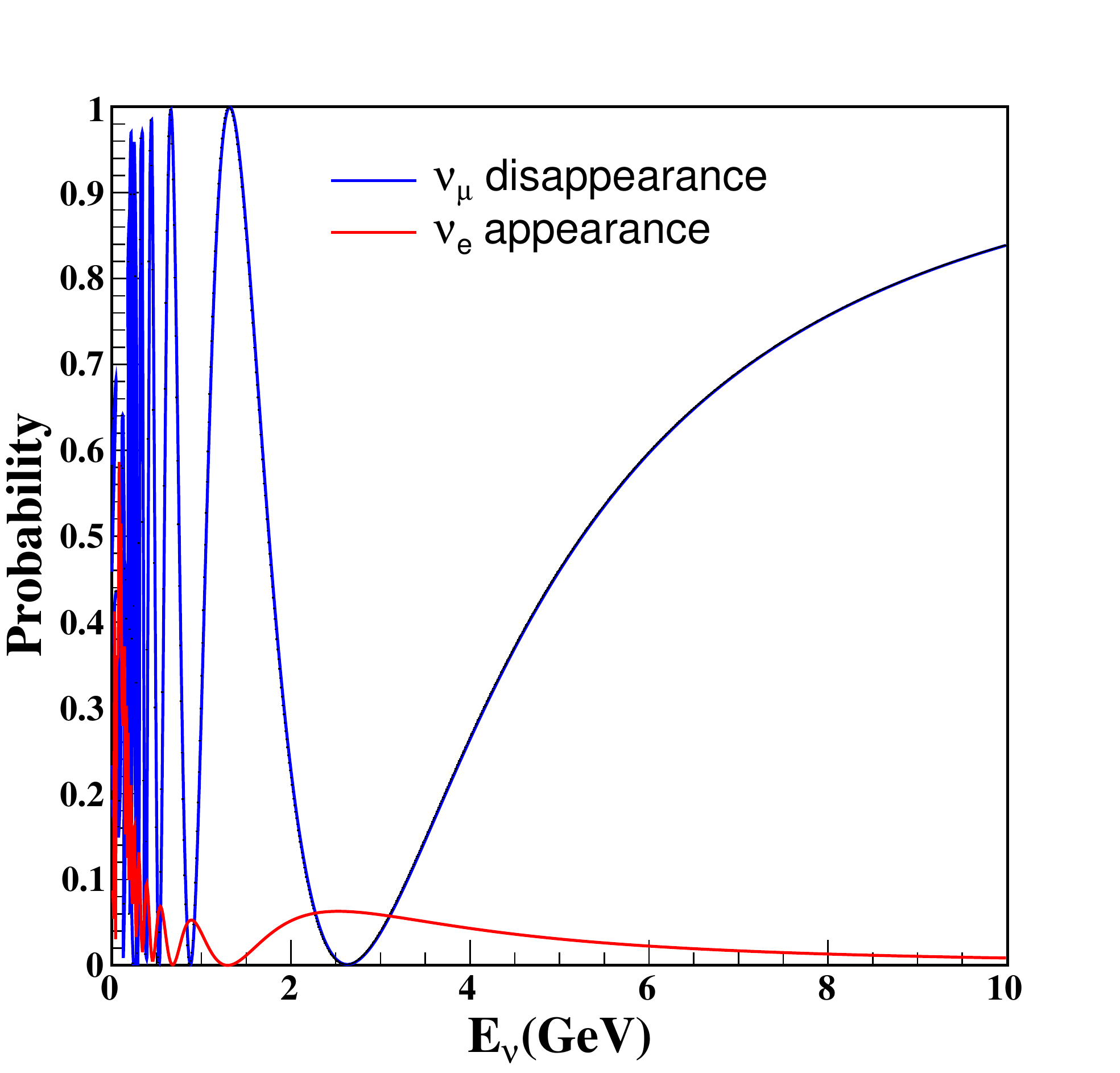}
\caption{Flux distribution in the energy regime 0.1-10 GeV and neutrino oscillation probability presented in the left and the right panels respectively.}
\end{figure}

\section{Final State Interactions and Nuclear Effects}
\begin{figure}
 \centering\includegraphics[scale=.5]{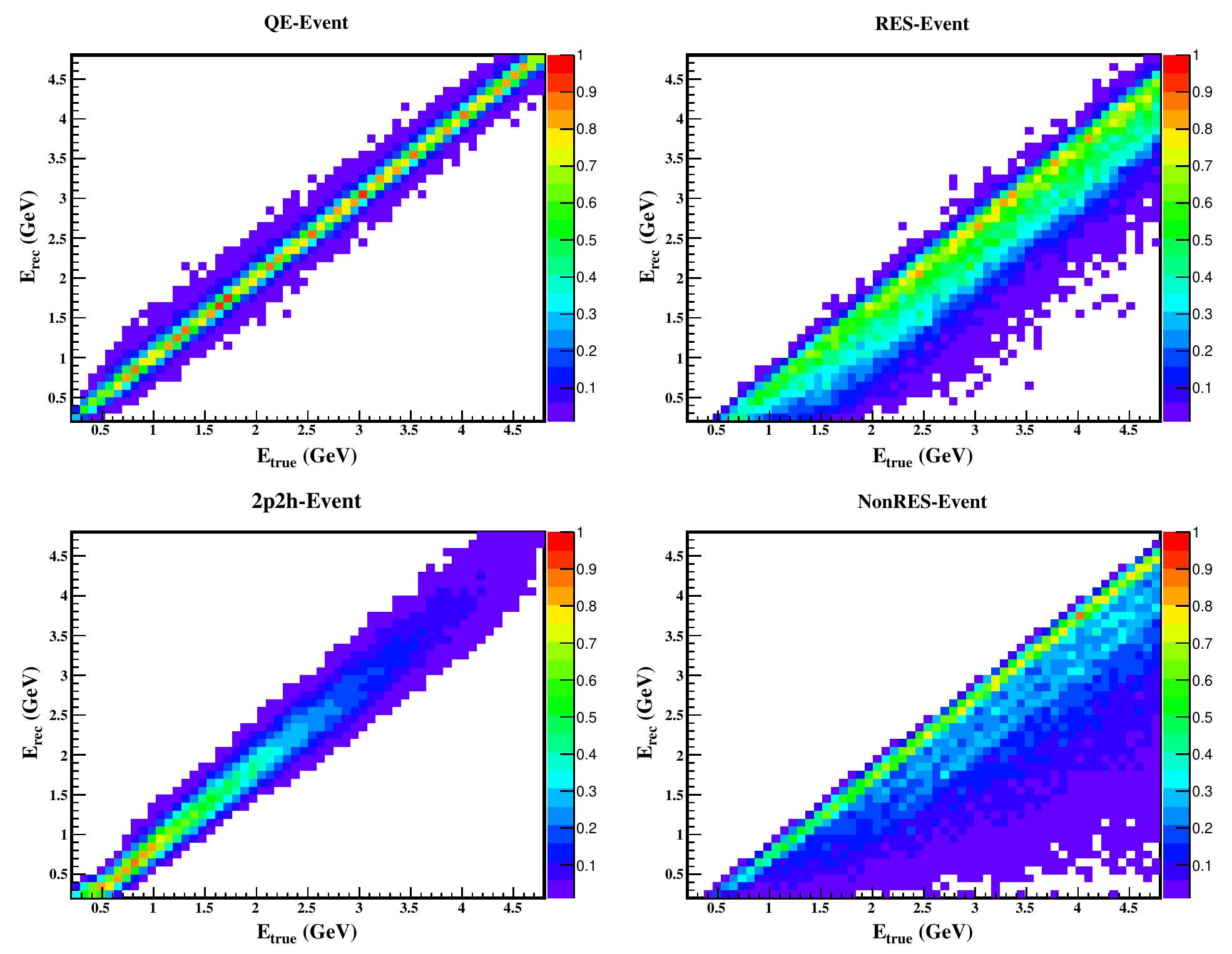}
 \caption{Two-Dimensional Migration matrices for QE, RES, 2p-2h and Non-RES Events(in top and bottom panels as indicated) generated using GiBUU for argon target.}
\end{figure}
A neutrino when interacts with a fermi moving neutron within the nucleus giving a muon and a proton as final state products is defined as a true charged current quasi elastic(CC-QE) process and is represented as : $\nu n \rightarrow \mu^{-}p$\\ 
These processes are accompanied by certain other processes in which a  pion or a delta resonance is also produced in the initial neutrino-nucleon interaction vertex  e.g. $\nu p \rightarrow \mu^{-} p \pi^{+}$ or $\nu p \rightarrow \mu^{-} \Delta^{++}$, but gets absorbed in the nuclear environment while propagating out of the nucleus via final state interactions. Such particles are not captured by the detector since they are not able to come out from the nucleus and thus lead to missing energy and appear as QE-like events. These events which are not produced as QE but appear as QE are known as fake-QE events (stuck pion event). These events are counted by the detector as true-QE events though they are not of true-QE origin and thus lead to uncertainties in the QE cross section measurement.
Contrary to the above process, another possibility is the production of a nucleon in the initial neutrino-nucleon QE interaction, which might get rescattered within the nucleus and produce a pion in the final state. Such an event, though having a true QE origin will be tagged as QE-like event due to the presence of a pion in the final state. The appearance of such events give way to uncertainties in the reconstruction of neutrino energy and the error caused in the energy reconstruction of neutrino, is further propagated to the measurement of neutrino oscillation parameters.
Figure 2 presents migration matrices constructed between true (t) and reconstructed (r) neutrino energies for QE, RES, 2p2h(multi-nucleon processes) and Non-RES interaction processes, using argon as the nuclear target. The migration matrices are computed following the technique mentioned in the ref \cite{coloma}. The neutrino-argon interaction events are generated in the energy range 0.2 to 5 GeV, further it is binned into 46 equal bins to observe the smearing in the reconstructed neutrino energy for different interaction processes as described above.
 
Every element n($E_{t},E_{r}$) in each of the migration matrices represents the probability that an event which is produced with a certain true energy $E_{t}$ is observed with some different energy $E_{r}$. Any difference in energy between $E_{t}$ and $E_{r}$ produces smearing along the diagonal line, which is evident from Figure 2. If the true and reconstructed neutrino energies are the same, then we will get a diagonal line without any smearing. The Non-RES and RES events appear to smear maximally while the 2p2h and QE events smear to a lesser extent as compared to the other two processes. From Figure 2, we can see that the error in energy reconstruction occurring in the various processes cannot be ignored if one has to aim toward reducing the systematic uncertainties in the forthcoming neutrino oscillation experiments.

\section{Simulation and Result}
Approximately 2$\times 10^{5}$ events are generated using DUNE flux for muon disappearance channel with the help of GiBUU(Giessen Boltzmann-Uehling-Uhlenbeck) \cite{gibuu,gibuu3}. The GiBUU model developed as a transport model for various interaction processes induced by nucleus, nucleon, pion and electron is based on a coupled set of semi-classical kinetic equations \cite{gibuu2}. GiBUU applies a semi-classical approach and models the FSI by solving the Boltzmann-Uehling-Uhlenbeck equations. Recently, the GiBUU code has also included the DIS interaction process successfully \cite{gibuuDIS}. The migration matrices are generated using GiBUU and are inserted in the required format into GLoBES \cite{globes1,globes2,globes3,globes4,globes5,globes6}. The systematics considered in our work are as follows- signal efficiency is 85$\%$, normalization error and energy calibration error for the signal and background are- 5$\%$, 10$\%$ and 2$\%$ respectively. The running time considered is 10 years in neutrino mode with 35 kton fiducial mass of the detector. The values of oscillation parameters used in this work are presented in Table 1 and are motivated from \cite{para1,para2}. The value of $\delta_{CP}$ considered in this work lie within the present set bounds on $\delta_{CP}$ \cite{globalrecent}.

\begin{table}[htp]
\caption{True Oscillation Parameters considered in our work \cite{para1,para2}.}
\renewcommand\thetable{\Roman{table}}
\centering
\setlength{\tabcolsep}{2pt}
\begin{tabular}{c | c  }
\hline
\hline
   $\theta_{12}$          &   $33.58^{\circ}$  \\  
   $\theta_{13}$          &   $8.48^{\circ}$  \\   
   $\theta_{23}$          &   $45^{\circ}$ \\
   $\delta_{CP}$          &   $180^{\circ}$  \\
   $\Delta m^{2}_{21}$    &   $7.50e^{-5}eV^{2}$ \\
   $\Delta m^{2}_{31}$    &   $2.40e^{-3}eV^{2}$  \\
\hline
\hline
\end{tabular}
\end{table}

\begin{figure}
\includegraphics[width=6.5cm, height=5.5cm]{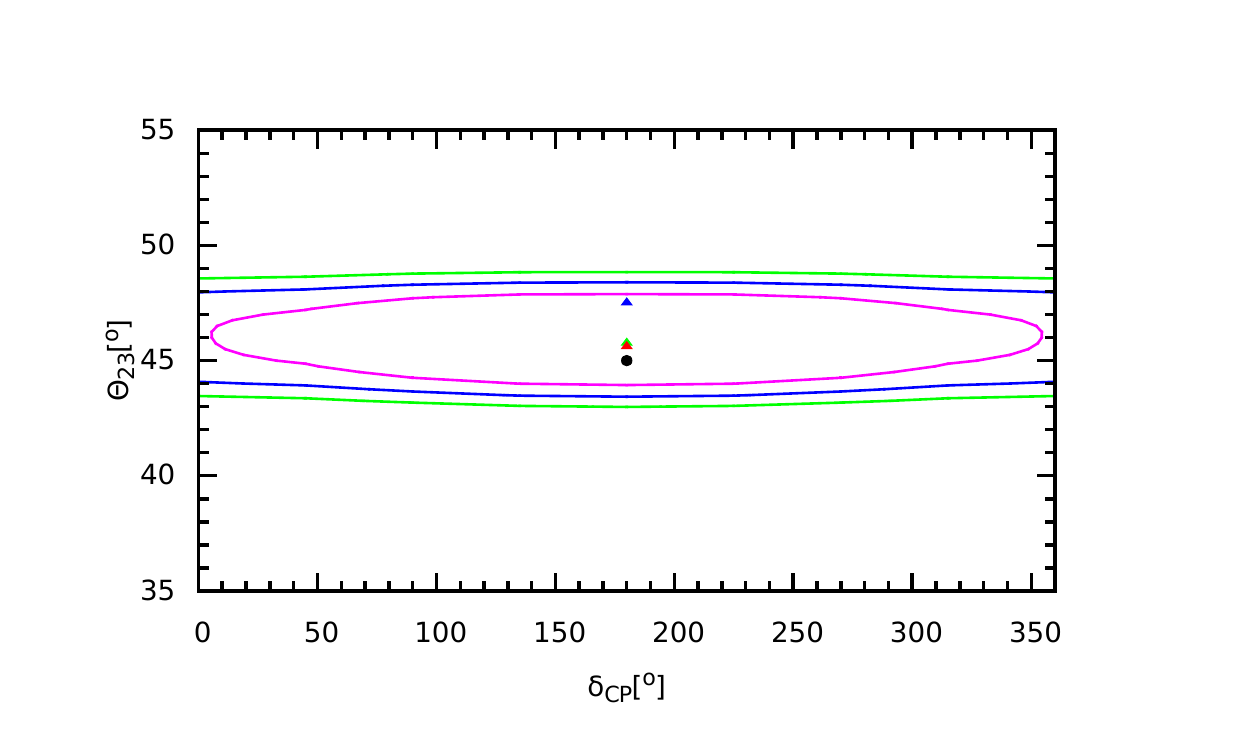} 
\includegraphics[width=6.5cm, height=5.5cm]{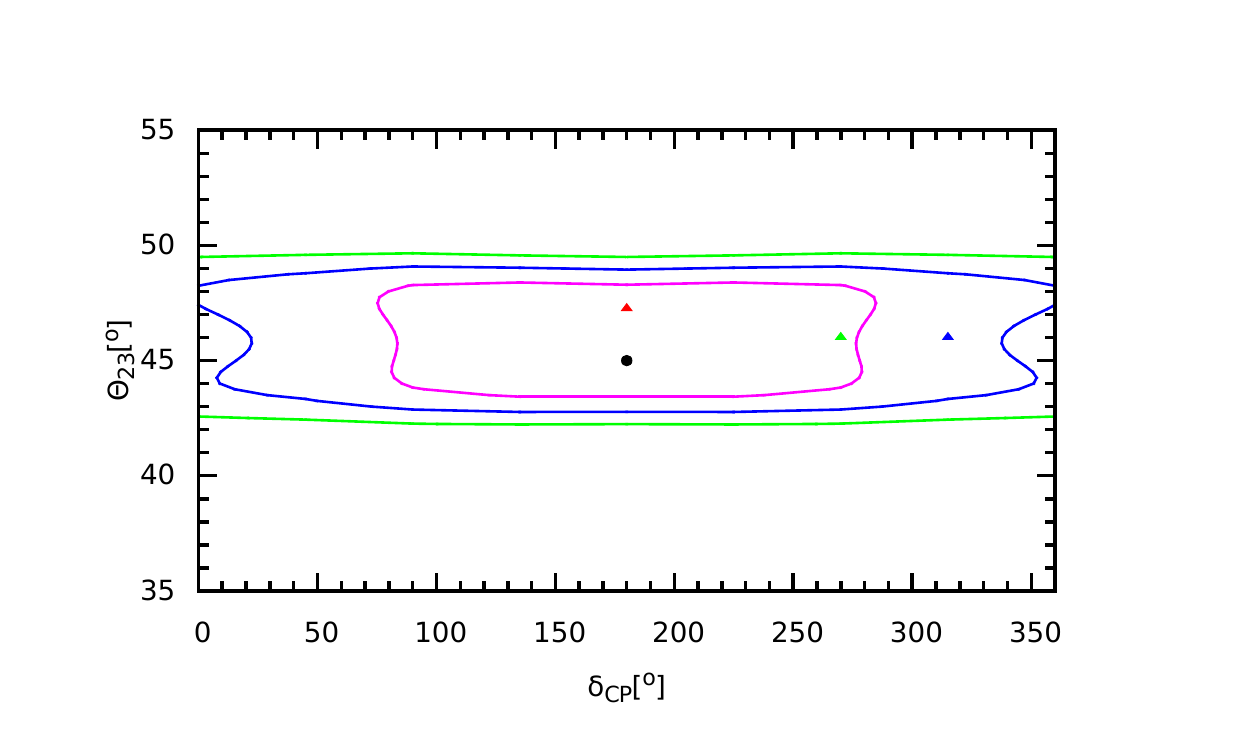}
\caption{Confidence regions in the($\theta_{23} ,\delta_{CP}$) plane, obtained by using the migration matrices of pure QE and QE like events are shown in the left panel and by using the migration matrices of pure QE+RES+DIS and QE+RES+DIS like events are shown in the right panel. The circle($\alpha$=1) shows the true value of the oscillation parameters while the blue($\alpha$=0), red($\alpha$=0.3) and green($\alpha$=0.5) triangles represent  the best fit points.}
\end{figure}

Here, we consider a parameter $\alpha$, which help us to incorporate nuclear effects in our analysis and explore the impact of nuclear effects in two extreme cases i.e. a case in which we have complete knowledge of nuclear effects and other case in which nuclear effects are completely neglected. It can be considered as a way of including systematic uncertainties, such an approach has been considered previously in \cite{coloma2,snaaz}.

We present the position of the best fit corresponding to values of $\alpha$ taken as 0 and 1, by plugging them in equations (1) and (2)-

\begin{equation}
N_{i}^{test}(\alpha) = \alpha \times  N_{i}^{QE} + (1-\alpha ) \times  N_{i}^{QE-like}
\end{equation}
\begin{equation}
N_{i}^{test}(\alpha) = \alpha \times N_{i}^{QE+RES+DIS} + (1-\alpha ) \times  N_{i}^{QE-like+RES-like+DIS-like} 
\end{equation}
where N is the total number of events. Here, $N_{i}^{QE}$ in equation (1) represents the pure QE events i.e. at the initial neutrino nucleon interaction vertex when the neutrino interacts with the nucleon the interaction was of QE nature. The products(lepton and a hadron) remain unchanged after traveling through the nucleus i.e. nature of the initial interaction remains unchanged and is considered as QE type interaction. Whereas, $N_{i}^{QE-like}$ are those events which at the initial neutrino nucleon interaction vertex were not of pure QE origin but as the products travel inside the nucleus their initial nature gets modified due to nuclear effects and at the time of detection they appear as QE type interaction.
 Similarly, $N_{i}^{QE+RES+DIS}$ in equation (2) is the number of events which originate from true QE, Resonance and DIS interaction processes while $N_{i}^{QE-like+RES-like+DIS-like}$ is the number of events which have a contribution from other processes as well.\\
Two cases arise-\\
1. When $\alpha$ = 1, the second term present in each of the equations (1) and (2) do not contribute to the event rate which implies that nuclear effects are completely neglected.\\
2. When $\alpha$ = 0, nuclear effects are perfectly known, the contribution arising from the first term of the equations (1) and (2) vanishes.

In the real scenario, our knowledge of nuclear effects lies between these two extreme cases. To quantify the error on the models defining nuclear effects, the introduction of the parameter $\alpha$ can help us to study the above two extreme cases. '$\alpha$' simply indicates the presence of an error occurring due to insufficient knowledge of nuclear effects.
We have also performed the analysis at $\delta_{CP}$=0$^{\circ}$ for the QE+RES+DIS event sample which can be found in appendix A1.



\enlargethispage{3\baselineskip}

\section{Conclusion}
Neutrino interactions on a heavy nuclear target(high atomic number i.e. A$>$12) are severely affected due to the presence of nuclear effects. We have performed our analysis for the Argon target which will be used in DUNE. The study is performed for the QE interaction channel and for a combination of QE, Res and DIS interaction processes. The impact of FSI on CCQE and CCRes interaction channels has been previously studied on atmospheric oscillation parameters in the refs \cite{coloma2, snaaz}. We have tried to explore the impact of FSI in $\theta_{23}-\delta_{CP}$ plane in both the cases (i) only QE event (ii) for QE+RES+DIS events.

The positions of the best fit point corresponding to four values of $\alpha \sim$ 0, 0.3, 0.5 and 1 for CCQE and CC(QE+RES+DIS) processes are represented in the left and right panels of Figure 3 respectively. The best fit points for $\alpha$=0.3, $\alpha$=0.5 and $\alpha$=1 are represented by red, green and blue triangles respectively. We notice that the deviation of the best fit point from the true input values increases as the value of $\alpha$ increases. In the left panel of Figure 3, we observe a slight shift in the best fit values of oscillation parameters from the true value($\theta_{23}$), if we increase the value of $\alpha$ from $\alpha$=0 to $\alpha$=0.5. The best fit values of $\alpha$=0.3 and $\alpha$=0.5 nearly overlap. This indicates that if the value of $\alpha$ is changed by a small amount(for $\alpha$ $<$ 0.5), a slight change is observed in the best fit value of oscillation parameters $\theta_{23}$ in QE case. But as we increase the value of $\alpha$ beyond 0.5, a remarkable change of $\sim$1 $\sigma$ is observed in the determination of $\theta_{23}$. If we take a data sample where events arising via QE+RES+DIS channels are taken into consideration, for this situation with $\alpha$=0.3, we observe a roughly $<$1$\sigma$ shift in the value of $\theta_{23}$ whereas if the value of $\alpha$ is increased to 0.5, there is a negligible shift in the true value of $\theta_{23}$ but a nearly $\sim$1$\sigma$ shift in the value of $\delta_{CP}$ is observed. As we increase the value of $\alpha >$ 0.5, the shift in the $\delta_{CP}$ value increases and gives a shift of roughly 2$\sigma$.



Our results show that for an experiment observing QE, Res and DIS events, with 50$\%$ knowledge of FSI there is a $\sim$ 1$\sigma$ bias in the extracted value of $\delta_{CP}$. If we have complete knowledge of FSI, the bias in the extracted value of $\delta_{CP}$ goes beyond 1$\sigma$. For an experiment observing only QE like events, a 1$\sigma$ bias in the determination of mixing angle $\theta_{23}$ is noted. These results indicate the impact of nuclear effects that may result from uncertainties present in the nuclear models.

In an outlook of the study, we can conclude that the best strategy for third-generation neutrino-oscillation experiments seems to minimize detection thresholds of the employed detectors and to perform an extensive authentication of the accuracy of nuclear models employed in data analysis. Employment of nuclear targets in neutrino oscillation experiments aid in boosting the event statistics which reduce the statistical error but we need to pin down the systematic uncertainties arising from the persistent nuclear effects that will bring us a step closer to achieving our goals.


\section{Acknowledgement}
This work is partially supported by the Department of Physics, Lucknow University. Financially it is supported by the government of India, DST project no-SR/MF/PS02/2013, Department of Physics, Lucknow.

\appendix

Here we present the confidence regions in the $\theta_{23}-\delta_{CP}$ plane using the migration matrices of pure QE+RES+DIS and QE+RES+DIS-like events with a different true input value of $\delta_{CP}$= 0$^{\circ}$. The positions of the best fit point corresponding to four values of $\alpha \sim$ 0, 0.3, 0.5 and 1 for CC(QE+RES+DIS) processes are represented in the Figure A1. We notice that the deviation of the best fit point from the true input value increases as the value of $\alpha$ increases. For this situation with $\alpha$=0.3, we observe a slight shift in the value of $\theta_{23}$ only whereas if the value of $\alpha$ is increased to 0.5, there is a negligible shift in the true value of $\theta_{23}$ but a nearly $\sim$2$\sigma$ shift in the value of $\delta_{CP}$ is observed. As we increase the value of $\alpha >$ 0.5, the shift in the $\delta_{CP}$ value increases and gives a shift of roughly 2$\sigma$. The nature of shift of the best fit point is similar to the one illustrated in right panel of Figure 3. From this, we can say that the impact of nuclear effects is not affected by the choice of the input parameters.

\begin{figure}
\includegraphics[width=6.5cm, height=5.5cm]{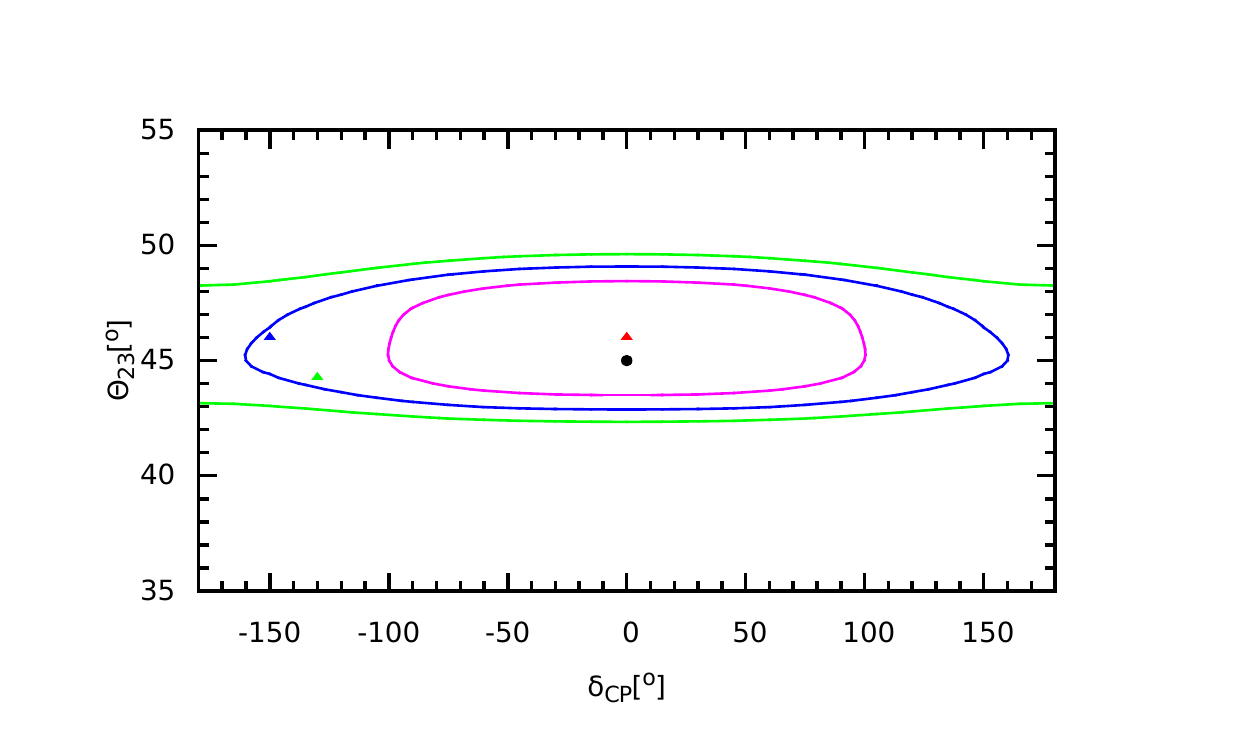}
\caption{Confidence regions in the($\theta_{23} ,\delta_{CP}$) plane, obtained by using the migration matrices of pure QE+RES+DIS and QE+RES+DIS like events. The circle($\alpha$=1) shows the true value of the oscillation parameters while the blue($\alpha$=0), red($\alpha$=0.3) and green($\alpha$=0.5) triangles represent  the best fit points.}
\end{figure}

\end{document}